\documentclass[fleqn,usenatbib,useAMS]{mnras}

\usepackage{graphicx}	
\usepackage{amsmath}	
\usepackage{amssymb}	
\usepackage{multicol}        
\usepackage{multirow}
\usepackage{bm}		
\usepackage{pdflscape}	

\usepackage{xspace}
\usepackage[capitalize]{cleveref}
\usepackage[dvipsnames]{xcolor}
\usepackage{siunitx}

\usepackage{subcaption}
\usepackage[T1]{fontenc}
\usepackage{ae,aecompl}
\usepackage{newtxtext,newtxmath}
\usepackage{ulem}
\usepackage{soul}
\pubyear{2022}

\begin{document}
\label{firstpage}
\pagerange{\pageref{firstpage}--\pageref{lastpage}}


\title[Vela X-1 observed by Insight-HXMT]{Variations of cyclotron resonant scattering features in Vela X-1 revealed with Insight-HXMT}
\author[Q. Liu et al.]{\bf \small Q. Liu$^{1,2}$, W. Wang$^{1,2}$\thanks{E-mail: wangwei2017@whu.edu.cn}, X. Chen$^{1,2}$, Y. Z. Ding$^{1,2}$, F. J. Lu$^{3}$, L. M. Song$^{3}$, J. L. Qu$^{3}$, S. Zhang$^{3}$, and S. N. Zhang$^{3}$ \\
$^{1}$School of Physics and Technology, Wuhan University, Wuhan 430072, China\\
$^{2}$WHU-NAOC Joint Center for Astronomy, Wuhan University, Wuhan 430072, China\\
$^{3}$Key Laboratory of Particle Astrophysics, Institute of High Energy Physics, Chinese Academy of Sciences, Beijing 100049, China
}
\maketitle
\begin{abstract}
We present a detailed study of the high mass X-ray binary Vela X-1, using observations performed by Insight-HXMT in 2019 and 2020, concentrating on timing analysis and spectral studies including pulse phase-resolved spectroscopy. The cyclotron line energy is found to be at $\sim 21-27$ keV and $43 -50$ keV for the fundamental and first harmonic, respectively. We present the evolution of spectral parameters and find that two line centroid energy ratio $E_2/E_1$ evolved from $\sim 2$ before MJD 58900 to $\sim 1.7$ after that. The harmonic cyclotron line energy has no relation to the luminosity but the fundamental line energy shows a positive correlation with X-ray luminosity, suggesting that Vela X-1 is located in the sub-critical accreting regime. In addition, the pulse phase-resolved spectroscopy in Vela X-1 is performed. Both the CRSF and continuum parameters show strong variability over the pulse phase with the ratio of two line energies about 2 near the peak phases, and down to $\sim$ 1.6 around off-peak phases. Long-term significant variations of the absorption column density and its evolution over the pulse phase may imply the existence of the clumpy wind structure near the neutron star.

\end{abstract}


\section{Introduction}
\label{sec:introduction}

Accretion-powered X-ray binary pulsars \citep{1989PASJ...41....1N,1997ApJS..113..367B} which consist
of a highly magnetized neutron star (NS) and an optical companion star, provide an individual laboratory for understanding the accretion process and properties of accretion geometry. For accretion onto the neutron stars, the falling gas from a donor is channeled to neutron star surface across the magnetic field , funneled to a relatively small region at the neutron star surface near the two magnetic polar caps, forming the hot spots and accretion columns where X-ray emissions occur \citep{1973ApJ...184..271L}.

Cyclotron resonant scattering features (CRSFs) are identified as characteristic absorption in the continuum spectrum of these X-ray pulsars \citep{2019A&A...622A..61S}. In NS, electrons perpendicular to the strong magnetic fields can be quantized into discrete Landau levels \citep{2007A&A...472..353S}, and photons with energies close to these levels can lead to cyclotron lines by scattering off these electrons in line-forming region. The magnetic fields strength on the surface of neutron stars can be directly measured by the cyclotron lines and calculated using
\begin{equation}
    \label{equ:B_En}
    B=\frac{E_{\rm cyc}(1+z)}{11.57\ \mathrm{keV}} 10^{12} \mathrm{G}
\end{equation}
where $E_{\rm cyc}$ is the centroid energy of the CRSF in units of keV \citep{1991ApJ...374..687H}, and $z$ is the gravitational redshift.

Vela X-1 is a wind-accreting ($\Dot{M} \sim 4 \times 10^{-6}  M_{\odot}/\rm yr$, \citealt{1986PASJ...38..547N}) and eclipsing high-mass neutron star X-ray binary consisting of a neutron star of $\sim$ 1.77$M_{\odot}$ \citep{2011ApJ...730...25R} whose rotation period is $\sim$ 283 s \citep{1976ApJ...206L..99M} with an 8.964 d \citep{1995A&A...303..483V} orbital period and an optical companion B0.5Ib super-giant HD 77523 \citep{1972ApJ...175L..19H}. The system is at a distance of 1.9 $\pm$ 0.2 kpc \citep{1985ApJ...288..284S} and its separation of the binary system is just 1.7 $R_{\star}$ \citep{2003A&A...401..313Q}. The companion has a mass of $\sim$23 $M_{\odot}$ and a radius of $\sim$30 $R_{\odot}$ with strong stellar winds \citep{1995A&A...303..483V}. Thus, the neutron star is deeply embedded in the dense stellar wind due to the extreme closeness, emitting X-rays by wind-fed accretion processes. The X-ray luminosity is typically about 5$\times 10^{36} $ erg s$^{-1}$ \citep{2010A&A...519A..37F}, and can sometimes show strong variability. 

The first detection for the cyclotron line near 50 keV in Vela X-1 was found by \cite{1992fxra.conf...51K}. Then \cite{1992fxra.conf...23M} and \cite{1996ApJ...471..447C} reported an absorption line feature at $\sim$32 keV with Ginga data. The existence of both two cyclotron line features around $\sim$23 keV and $\sim$45 keV was reported by \cite{1997A&A...325..623K} using the data from HEXE. \cite{1998A&A...332..121O} used the BeppoSAX data to study Vela X-1, but only confirmed the line at $\sim 55$ keV. \cite{2002A&A...395..129K} using RXTE data confirmed the existence of the two lines at $\sim$ 25 and 55 keV. In recent years, \cite{2013ApJ...763...79M} reported the detection of CRSFs at $\sim$25 and $\sim$50 keV with Suzaku data. And \cite{2014MNRAS.440.1114W} also confirmed the CRSFs at about 25 and 55 keV in Vela X-1 by INTEGRAL. Many researchers investigated the correlation between CRSFs and flux in accreting X-ray pulsars \citep{2012A&A...542L..28K}. \cite{2014ApJ...780..133F} showed the anti-correlation between the depths of the two cyclotron lines in Vela X-1 and the positive relation between the first harmonic line energy and X-ray luminosity. These studies can help to understand the accretion geometry and physical processes onto the surface of the strongly magnetized neutron stars.

Furthermore, the CRSF parameters over different pulse phases can provide vital information on the magnetic field structure of the emission region of the neutron star. \cite{phd} reported that the CRSFs of Vela X-1 varied in strength in different pulse phases with Ginga data. \cite{1999A&A...341..141K} confirmed the existence of the two lines with the pulse phase using Rossi X-ray Timing Explorer (RXTE) data. \cite{2003A&A...400..993L} also confirmed the presence of the cyclotron absorption line at $\sim$55 keV in Vela X-1 with phase-resolved analysis based on BeppoSAX data. \cite{2013ApJ...763...79M} presented a pulse-phase-resolved spectral analysis using Suzaku observations during 2008, and found both two line parameters varying with pulse phase.

In this paper, we studied the properties of Vela X-1 and two cyclotron lines based on the hard X-ray spectra ranging from 2 to 105 keV, and analyzed their variations and correlations in phase averaged spectrum using 2019 and 2020 data with Insight-HXMT. To study spectral properties at different viewing angles, we also carried out phase-resolved spectroscopy. The paper is structured as follows. In Section 2, we briefly describe the HXMT observations and data reduction. Timing analysis is given in Section 3. The phase averaged and resolved X-ray spectral studies of Vela X-1 are both included in Section 4. The results and possible implications of the variations and correlations of CRSFs are discussed in Section 5. The conclusion is summarized in Section 6.

\section{Insight-HXMT Observations and Data reduction} \label{sec:optimized}

The Hard X-ray Modulation Telescope (HXMT) \citep{2020SCPMA..6349502Z} launched on 15th June 2017, is the first X-ray astronomical satellite in China, which consists of three instruments with a total energy band of 1–250 keV : High Energy X-ray telescope (HE) from 20 to 250 keV band with detector areas of $\sim$5000 $\rm cm^2$, Medium Energy X-ray telescope (ME) for 5–30 keV band having areas of 952 $\rm cm^2$, and Low Energy X-ray telescope (LE) for 1–15 keV band with a total area of 384 $\rm cm^2$.

The HE contains 18 NaI(Tl)/CsI(Na) scintillation detectors, which include 15 narrow FOV($5.7^\circ \times 1.1^\circ$), 2 wide FOV($5.7^\circ \times 5.7^\circ $) and a blind FOV. Both of ME and LE X-ray telescopes consist of three detector boxes. Each detector box of ME has 576 Si-PIN detector pixels. Each detector box of LE contains 32 CCD236 (a kind of Swept Charge Devices). The time resolution of detector for HE, ME, and LE is 25 $\mathrm{\mu} s$ , 276 $\rm \mu s$ and 1 ms, respectively.  

The Insight-HXMT Data Analysis Software (HXMTDAS) v2.04 is used for data analysis. To calibrate and screen events for three payloads, we require good time intervals (GTIs) in HXMTDAS using hegtigen, megtigen, and legtigen tasks which all adopt the pointing offset angle $<0.04^\circ$, the elevation angle $>10^\circ$, and the geomagnetic cut-off rigidity >8 GeV. Background light curves are estimated by using hebkgmap, mebkgmap, and lebkgmap tasks. Tasks helcgen, melcgen, and lelcgen are used to extract X-ray light curves with $\sim$ 0.008 (1/128) sec time bins.

The ID lists and related information of the utilized observations on Vela X-1 by Insight-HXMT are summarized in \cref{tab:ObsIDs}, including 5 observations in 2019 and 6 observations in 2020. For each single observation ID, Insight-HXMT also split each observation into multiple segments (called "exposure") that are identified as the Exposure IDs, and so, finally the observation data of total 28 exposure IDs are available for the timing and spectral analysis (see \cref{tab:out_fraction}). In order to perform the timing analysis, the arrival times of photons were corrected from terrestrial time to the solar system barycentric time at first. For spectral studies, we used XSPEC 12.11.1 version \citep{1996ASPC..101...17A} with several custom models for the spectral fitting analysis.

\begin{table}
    \centering
    \caption{The observation IDs of Insight-HXMT used in this study.}
    \label{tab:ObsIDs}
    \begin{tabular}{l|cccc}
    \hline \hline 
    \multirow{2}{*}{ObsID} & \multirow{2}{*}{Time Start [UTC]}  & \multirow{2}{*}{Duration [s]} & \multirow{2}{*}{MJD [D]}\\ \\
    \hline 
    P0101247002 & 2019-01-09T08:32:53 & 81313 & 58492\\
    P0101247003 & 2019-01-12T00:11:29 & 57666 & 58495 \\
    P0201012141 & 2019-11-02T06:19:13 & 166381 & 58790\\
    P0201012143 & 2019-12-15T22:44:29 & 41004 & 58833\\
    P0201012144 & 2019-12-16T19:25:08 & 104109 & 58834\\
    P0201012145 & 2020-01-03T04:11:05 & 138577 & 58852\\
    P0201012147 & 2020-01-22T03:15:58 & 172572 & 58871\\
    P0201012149 & 2020-03-22T09:10:47 & 172341 & 58932\\
    P0201012150 & 2020-04-20T22:39:52 & 172456 & 58960 \\
    P0201012151 & 2020-06-17T01:52:17 & 86443 & 59017\\
    P0201012152 & 2020-06-23T05:47:49 & 86453 & 59023\\
    \hline
    \end{tabular} 
\end{table}

\section{Timing analysis}

\begin{figure}
    \centering
    \includegraphics[width=.5\textwidth]{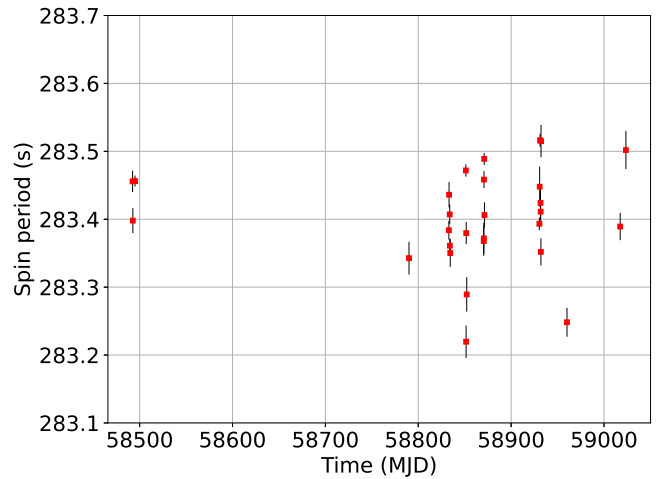}
    \caption{The intrinsic spin period evolution of the neutron star in Vela X-1 versus time.}
    \label{fig:period}
\end{figure}

We searched for the pulsed period by using the epoch folding method  \citep{1987A&A...180..275L} using the X-ray light curves of 0.008-sec bin. The Doppler effect resulting from the binary orbit motion was considered to modify the pulse frequency. The uncertainties of the spin period are estimated using a Gaussian error. As seen in \cref{fig:period}, the derived intrinsic pulse period ($\sim 283.4$ s) of the neutron star in Vela X-1 shows random variations due to unstable accretion wind. The change of the neutron star spin period shows no long-term evolution, and the fluctuation can be up to a fraction of 0.1\% in these observations. The energy resolved pulse profiles were finally created by folding the light curves in different energy bands with the obtained pulse period for each exposure ID.

\begin{figure}
    \centering
    \includegraphics[width=.5\textwidth]{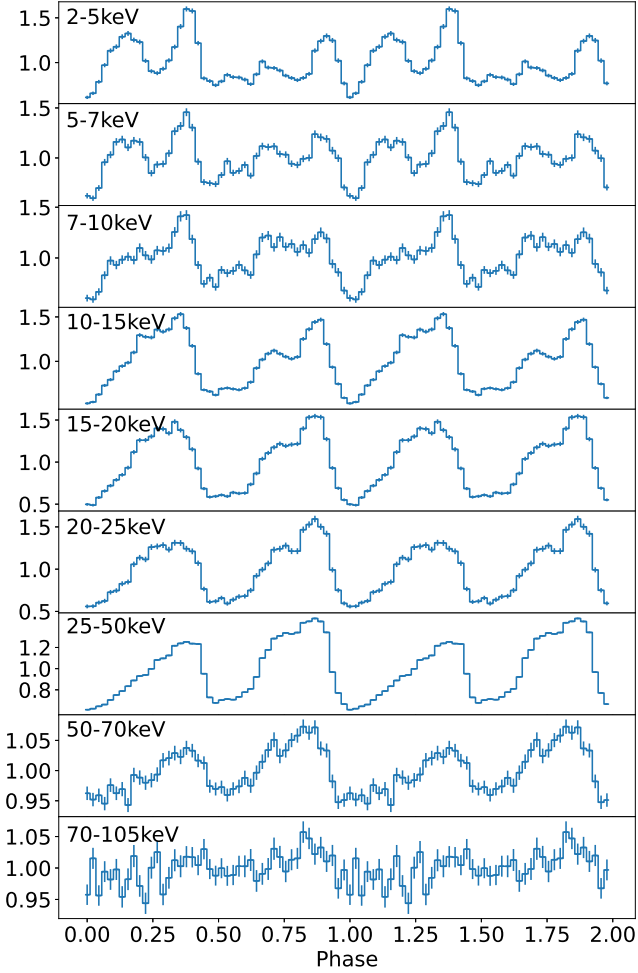}
    \caption{The pulse profiles of Vela X-1 for the ExpID P020101214901 (MJD-58930) in different energy ranges.}
    \label{fig:pulse}
\end{figure}

The pulse profile for the accreting pulsar is usually highly energy dependent (\citealt{2012MNRAS.423.2854L,2021MNRAS.506.2712D,2021JHEAp..30....1W}). As an example of Vela X-1, \cref{fig:pulse} presents the pulse profiles over various energy bands for one exposure ID. Above 7 keV, the pulse profile consists of two main peaks, and here we designate the 0-0.5 phase as the secondary pulse and the 0.5-1.0 phase as the main pulse. The double-peak structure in high energy bands is quite stable and has no evident evolution with time sequence. While the peak height of the secondary pulse is lower than that of the main pulse below $\sim 15$ keV, and above 15 keV, the main pulse becomes the brighter one. Above 70 keV, the pulse profiles are not obviously evident due to low detection significance levels.

Below 7 keV, these two pulses evolve into a complex multiple peaked structure which is consistent with the behavior found by \cite{1999A&A...341..141K}. As seen in \cref{fig:pulse}, the main pulse transits to complex three mini peaks and the secondary one changes to two mini peaks. 

Distinct differences exist in pulse profiles as the energy increases, with multiple peaks at low energies and double peaks at higher energies. The high energy double pulse profile above 10 keV can be ascribed to the magnetic dipole structure. However, there is currently no good explanation for the complex low energy pulse profiles \citep{1990A&A...234..172R}.

\section{Spectral Analysis} \label{sec:result}

\begin{figure*}
\centering
\includegraphics[width=1.0\textwidth]{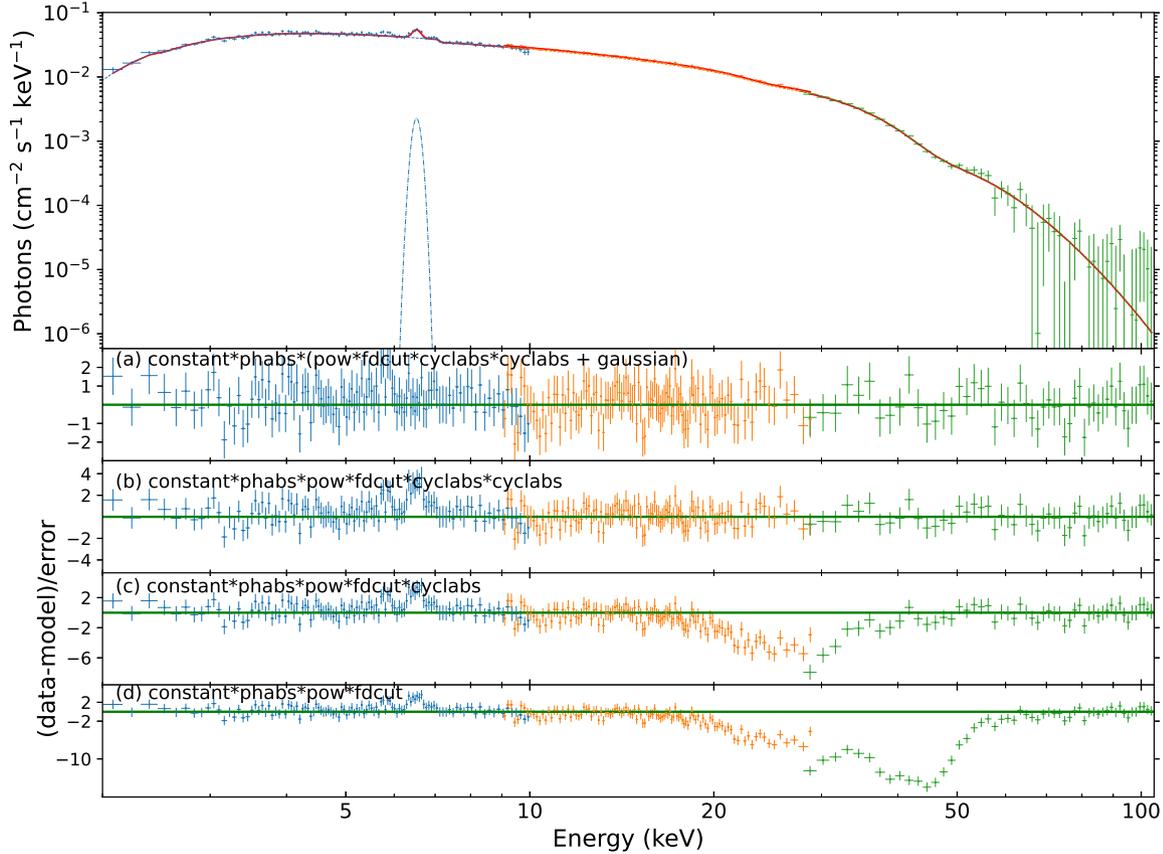}
\caption{The hard X-ray spectrum from 2–105 keV of Vela X-1 obtained by Insight-HXMT (an example based on Obs ID P02010124906) with the model fitting with the power-law with a FD-cut multiplied by phabs (XSPEC) plus two cyclotron scattering lines at $\sim$ 25 and 46 keV with an Fe K$\alpha$ line at $\sim$ 6.4 keV. The residuals after the spectral fitting are shown in the followings: (a) The spectrum is fitted with a FD-cut multiplied by phabs plus two cyclotron scattering lines with an Fe K$\alpha$ line. (b) The spectrum is fitted with a FD-cut multiplied by phabs plus two cyclotron scattering lines. (c) The spectrum is fitted with a FD-cut multiplied by phabs plus a cyclotron scattering line around 46 keV. (d) The spectrum is fitted with a FD-cut multiplied by phabs model.}
\label{fig:pp_0S1Q3}
\end{figure*}

\subsection{Phase averaged spectral analysis}

The shape of a continuum spectrum for X-ray pulsars is believed to be dominated by the emission of Comptonization of thermal photons \citep{1994A&A...284..161S, 1996ApJ...459..666A}. Since there are still no convincing theoretical models for the shape of X-ray continuum in accreting X-ray pulsars \citep{1994AIPC..308..429H}, the empirical models, usually a power-law component at low energies with an exponential cutoff at higher energies \citep{1983ApJ...270..711W, 1986LNP...255..198T} are chosen. To model the continuum spectra of Vela X-1, a simple power-law with high energy exponential roll-off (cutoffpl, XSPEC) does not fit the data. The Negative Positive Exponential (NPEX) model can provide an acceptable fit, but the fitting in the lower energy ($< 6$ keV) is not always good. The X-ray continuum is well described by a power-law model ($\propto E^{-\Gamma}$, where $\Gamma$ is the photon index) with the Fermi-Dirac cut-off (FDcut), which is commonly applied to the spectral fitting of Vela X-1 \citep{1999A&A...341..141K, 2014ApJ...780..133F} and has the following expression
\begin{equation}
    FDcut(E) \propto \left(1+\exp \left(\frac{E-E_{\mathrm{cut}}}{E_{\mathrm{fold}}}\right)\right)^{-1},
\end{equation}
where $E_{\rm cut}$ and $E_{\rm fold}$ are the cutoff and folding energies, respectively. Additionally, in order to describe the neutral hydrogen column absorption, we adopt phabs (XSPEC) model. To allow for the cross-calibration difference to be below 5\% in different payloads, we include a constant in the models. In \cref{fig:pp_0S1Q3}, we present the 2–105 keV spectrum fitting of the data based on Obs ID P02010124906 as an example. The fitting residual of $constant*phabs*pow*FDcut$ model is shown in panel (d) of \cref{fig:pp_0S1Q3}.

The fitting results of the above model show that obvious absorption line features exist in our data (see the residuals in panel (d)). In order to model the cyclotron absorption line component, we used a Lorentzian profile model (cyclabs in XSPEC) which is described with  
\begin{equation}
    CRSF(E)=\exp \left[-D_{\mathrm{f}} \frac{\left(W_{\mathrm{f}} E / E_{\mathrm{cyc}}\right)^{2}}{\left(E-E_{\mathrm{cyc}}\right)^{2}+W_{\mathrm{f}}^{2}}\right],
\end{equation}
where $D_{\mathrm{f}}$ represents the depth of the line, $W_{\mathrm{f}}$ describes the width of the line. As mentioned earlier, both CRSFs have been reported in the previous study, but since the fundamental CRSF is weak or shallow, sometimes only the harmonic can be detected. Considering this, we only fit the first harmonic line around 50 keV at first, and notice the absorption feature around 20 -- 35 keV still appears (refer to the residuals in panel (c)). Thus, the fundamental line is needed for the model fittings, which means both two CRSFs can be confirmed by Insight-HXMT data (see panel (b) in \cref{fig:pp_0S1Q3}).

Vela X-1 shows fluorescence lines associated with Fe K$\alpha$ or Fe K$\beta$ in its soft X-ray spectrum. \cite{2006ApJ...651..421W} reported this feature around $\sim$6.4 keV, and it can also be noted in the panel (b) of the residuals in \cref{fig:pp_0S1Q3}. Thus a Gaussian function around 6.4 keV was added in our model, although sometimes the Fe K$\alpha$ line in Vela X-1 can be hardly detected by Insight-HXMT due to lower resolution. It should be noted that the so-called 10 keV broad absorption feature in Vela X-1 reported by the NuStar observations (e.g. \citealt{2014ApJ...780..133F,2022arXiv220104169D}) cannot be confirmed by the Insight-HXMT data. Thus, in the following spectral fitting, we only consider the Fe K$\alpha$ emission feature in the band of 6 -- 10 keV. 

Therefore the final model for the spectral fittings in Vela X-1 from 2 -- 105 keV can be summarized as
\begin{equation}
   \begin{aligned}
   I(E) =  & Constant * N_\mathrm{H} *(pow*FDcut(E) \\ & 
    * CRSF(E,F) * CRSF(E,H) + Fe K \alpha),
   \end{aligned}
\end{equation}
where "CRSF(E,F)" and "CRSF(E,H)" represent the fundamental and
the harmonic line, respectively. The residuals of optimal fitting models can be seen in panel (a). In \cref{tab:out_fraction}, the best fitting parameters of all the observations are presented together.

In addition, we also adopted the Monte Carlo Markov Chain method with each chain of 50 000 steps to estimate the uncertainties of spectral parameters and check the degeneracy for some vital parameters. There is no apparent degeneration between these parameters, and we can briefly conclude that the first harmonic line in Vela X-1 is determined to be at the mean centroid energy of $\sim$ 46 keV and the fundamental line at the mean centroid energy of $\sim$ 25 keV, even though the fundamental lines are not always visible. Comparing the two cyclotron line features, it is obvious that the fundamental line is always much shallower than that of the first harmonic in Vela X-1 (see \cref{tab:out_fraction}), this is the possible reason that many previous observations could not confirm it \citep{1996ApJ...471..447C,1998A&A...332..121O,2003A&A...400..993L}.

\subsection{Phase-resolved spectroscopy analysis}

The spectra of accretion X-ray pulsars have been found to vary in different spin phases. These variations can be used to infer the accretion process of the emission region and geometric characteristics. We carried out a detailed pulse phase resolved spectral analysis of Vela X-1. For clarity, we choose the representative exposure numbers P020101214906, P020101214903, and P020101214914 due to higher detection significance of cyclotron lines. After the correction of barycentering and orbital motion of Vela X-1 by using the ephemeris \citep{1989PASJ...41....1N}, we divide the phase into 16 overlapping bins based on the derived pulse periods. For every phase resolved spectrum we regenerate the new background spectra and response matrices using the HXMT softwares. 

The phase resolved spectra at different phase bins can be well described by the same model used in the phase averaged case, and the fitting reduced $\chi^2$ statistical values are close to 1. However, due to the limited resolutions, the CRSFs (especially the fundamental line) sometimes are unable to be detected for some phases.

\section{Results and discussion}

\subsection{The variations of continuum and CRSF parameters}

\begin{figure}
\centering
\includegraphics[width=.5\textwidth]{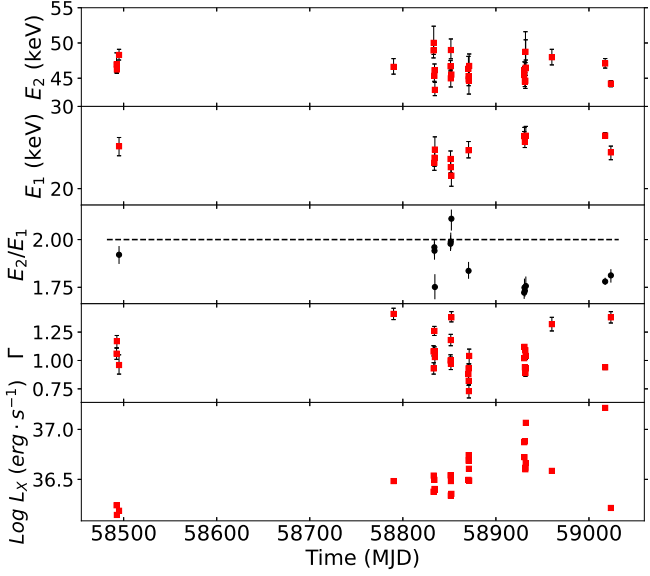}
\caption{The evolution of two cyclotron line energies, photon index and X-ray luminosity, the ratio of two line energies ($E_2/E_1$) versus time.}
\label{fig:Etime}
\end{figure}

The existence of long-term change in CRSF energy has been reported in some X-ray pulsars. \cite{2014A&A...572A.119S, 2016A&A...590A..91S} reported in Her X-1 the fundamental CRSF energy reduced by 5 keV from 1996 to 2015 with RXTE, Beppo-SAX, INTEGRAL, Suzaku, and NuSTAR, but the 20 years' decrease stopped and had an inverse trend \citep{2017A&A...606L..13S}. In Vela X-1, \cite{2010A&A...510A..47S} claimed that the first harmonic line energy decreased by $\sim$0.72 keV per year from 2004 to 2010 using Swift/BAT data and then remained constant after 2011. \cite{2016MNRAS.463..185L} also found that the CRSF harmonic energy of Vela X-1 decreased by $\sim$ 0.36 keV yr$^{-1}$ between 2004 -- 2013 based on Swift/BAT monitoring. Based on the Insight-HXMT results, we investigated the evolution of both two cyclotron line energies with time displayed in \cref{fig:Etime}. During a period from 2019 to 2020, we found that the first harmonic line energy fluctuating between $\sim$ 43 and $\sim$ 50 keV is quite stable which is still consistent with the previous work \citep{2010A&A...510A..47S}. The fundamental line energy of Vela X-1 is detected at 21-27 keV and has large variations but seemly no evolution with time. 

We also plot the ratio of two lines energies versus time, the ratio of $E_2/E_1$ was around 2 from MJD 58500 to MJD 58900, and $\sim 1.7$ after MJD 58900. \cite{2002A&A...395..129K} has reported a line energy ratio of $2.15 \pm 0.19$ of Vela X-1 in phase averaged spectrum with RXTE. Other sources such as 4U 0115+63 \citep{1999ApJ...521L..49H,1999ApJ...523L..85S,2012MNRAS.423.2854L} have shown that the fundamental line profile and centroid energy can intensely change and therefore the $E_2/E_1$ ratio can not be stable during the outburst. In addition, $\Gamma$ does not change with time, while the observed luminosity $L_{\rm X}$ has an increasing trend with time during the Insight-HXMT observations.


\begin{figure}
\centering
\includegraphics[width=.5\textwidth]{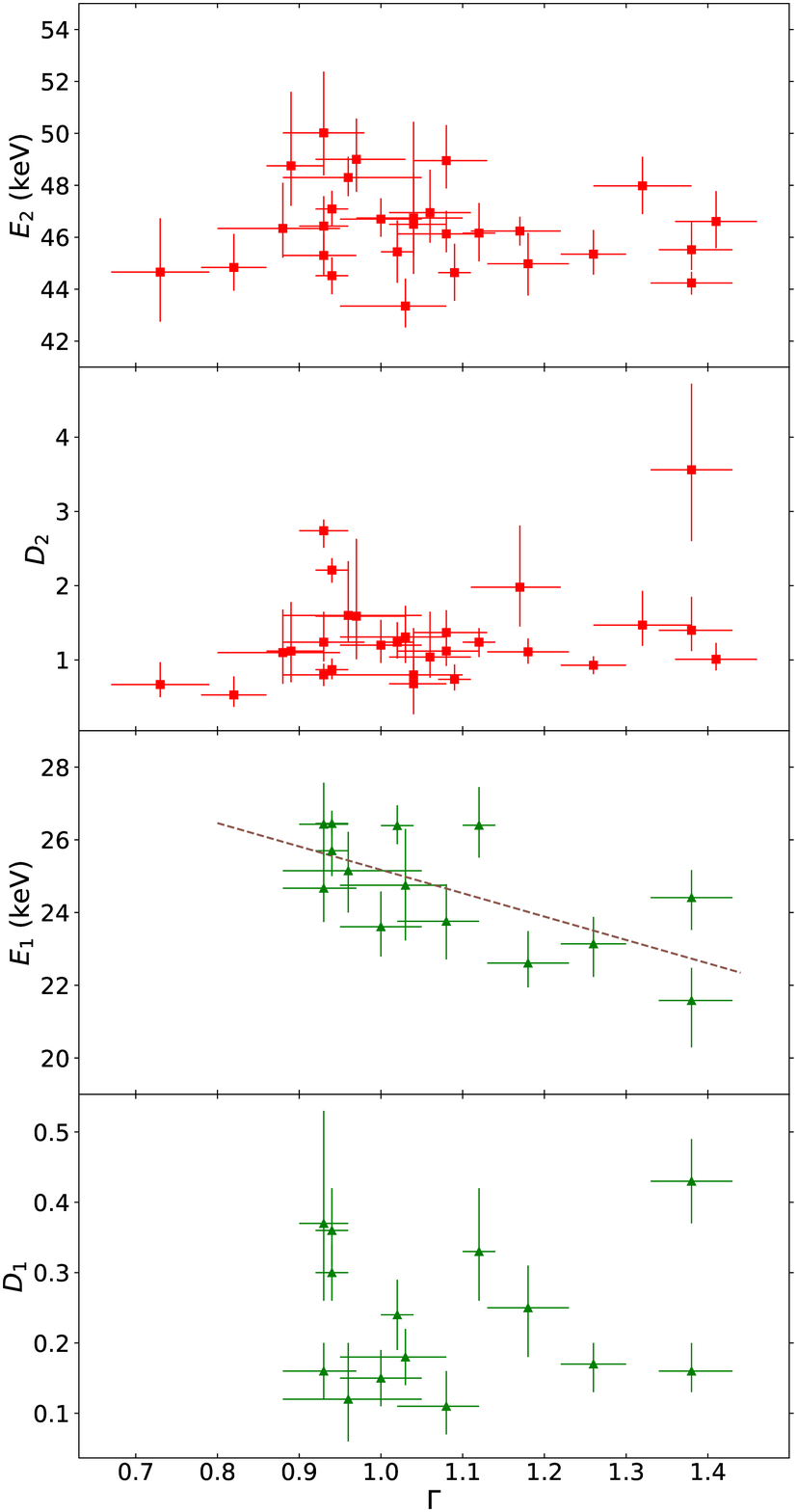}
\caption{The fundamental CRSF line energy $E_1$, depth $D_1$ and harmonic line energy $E_2$, depth $D_2$ as a function of photon index $\Gamma$. Red points show the cases of the harmonic line, green points represent the cases of the fundamental line. In the middle panel, the dashed line shows the best fitted function.}
\label{fig:E1E2D1D2_index}
\end{figure}

The energies and depths of two CRSF lines versus the photon index are shown in \cref{fig:E1E2D1D2_index}. The energy ($E_2$) and depth ($D_2$) of the harmonic line have no relation to $\Gamma$. However, the fundamental line energy shows a weak anti-correlation with the photon index, with a Pearson's correlation coefficient $r=-0.66$, with a best-fit slope of $ -7.8\pm 2.3$. 


\subsection{Luminosity dependence of the cyclotron line centroid energies}

\begin{figure}
\centering
\includegraphics[width=.5\textwidth]{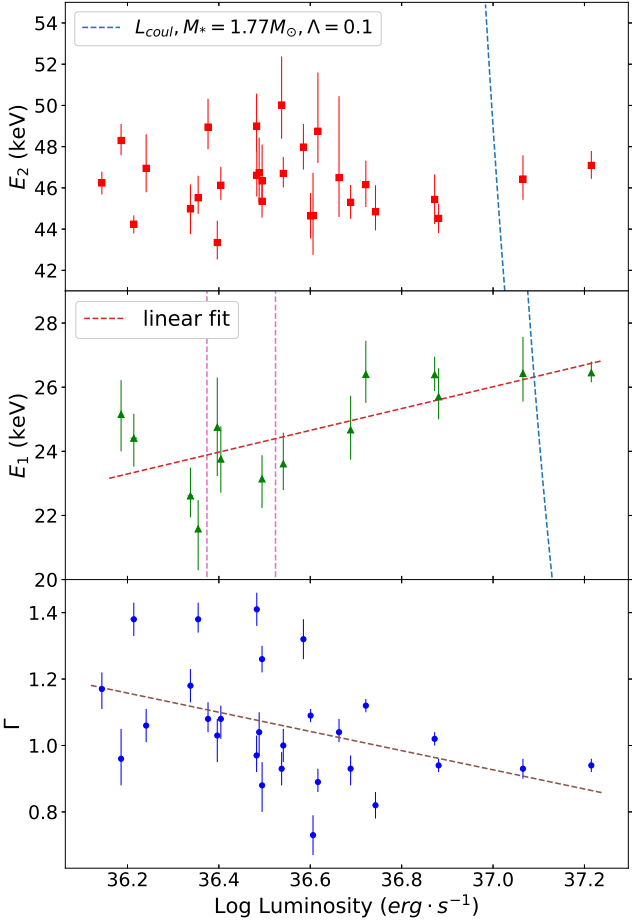}
\caption{The cyclotron line centroid energies as a function of X-ray luminosity: the upper panel shows the first harmonic line energy (red point), the middle panel shows the fundamental line (green point). The blue dashed line is the theoretical curve of $L_{\rm coul}$ using $M_{*} =1.77 M_{\odot}, \Lambda = 0.1$. The two vertical magenta dashed lines of the middle panel denote the curve of $L_{\rm coul}$ for the fundamental cyclotron energy $E_{1} = 25 \rm keV$ in the case of $\Lambda = 1$ with the neutron star mass $M_{*} =1.4 M_{\odot}$ in lower limit, and $M_{*} =1.8 M_{\odot}$ in higher limit respectively. The bottom panel also represents the photon index $\Gamma$ versus luminosity, the dashed line as the fitted linear function.}
\label{fig:GE1E2_Flux}
\end{figure}

The dependence of CRSFs on luminosity is important and probably reflects neutron stars' accretion regimes and geometry structure. In X-ray binary pulsars (XRBPs), the X-ray emission is usually expected to be produced in accretion columns \citep{1976MNRAS.175..395B}. There mainly exists two accretion dynamic regimes, i.e. the super-critical and sub-critical cases. Above the so-called critical luminosity $L_{\rm crit}$, the radiation-dominated shock whose emission height increases with luminosity decelerates the matter accretion. Below the critical luminosity, the Coulomb interaction breaks the infalling materiel and the emission height reduces with luminosity. Based on this assumption, the anti-correlation between CRSF energy and luminosity is expected in the super-critical regime, and a positive correlation is expected in the sub-critical regime. \cite{2014ApJ...780..133F} reported the positive correlation of the harmonic line energy with X-ray luminosity in Vela X-1 and suggested that the source is located in the sub-critical accretion regime, implying $L_{\rm crit}> 10^{37}\rm erg\ s^{-1}$.



In \cref{fig:GE1E2_Flux}, we presented the relationships between two CRSF line energies versus X-ray luminosity with Insight-HXMT data. The first harmonic line energy ranging from $\sim 43- 50$ keV has no visible relations with luminosity. For the case of the fundamental line, the centroid energy seems to show an evolution with the observed X-ray luminosity $L_\mathrm{X}$. At first, we use a linear function to fit the data points for the whole luminosity range, there exits a weak positive relationship between the line energy and luminosity with the Pearson's correlation coefficient $r\sim 0.69$ with a slope of $3.3\pm 0.7$ and a probability of $\sim 3\times 10^{-4}$, a reduced $\chi^2$ of 1.52 (12 degrees of freedom, hereafter d.o.f). This correlation is not so strong maybe due to the scattering of data points in the low luminosity range. The uncertainties also exist for NuStar observations \citep{2014ApJ...780..133F} and the other high mass X-ray pulsar GRO J1008-57 \citep{Chen_2021}. Then, we will test the relation of $E_1 - L_{\rm X}$ in two luminosity ranges. In the larger luminosity range of $> 2 \times 10 ^{36}$ erg s$^{-1}$, the linear function fit still gives a positive correlation with $r\sim 0.82$, a slope of $3.4\pm 0.8$, a probability of $\sim 0.002$ and reduced $\chi^2$ of 1.23 (8 d.o.f), then the positive correlation is not improved so much when we only use data points of the high luminosities. Below the luminosity, we also use a linear function to fit the data points, which gives $r\sim - 0.99$, with a slope of $-16.1\pm 2.5$ and a probability of $\sim 0.02$. Since there are only four data points, the negative relation could not be claimed significantly yet and need more observations in future. Thus the fundamental line energy shows the positive relation with X-ray luminosity based on Insight-HXMT, suggesting a sub-critical luminosity for the case of Vela X-1, which is consistent with the results of \cite{2014ApJ...780..133F}.

The positive relation could be consistent with the hypothesis that Coulomb interactions dominate shock deceleration of the gas in the sub-critical regime, where a coulomb luminosity $L_{\text {coul}}$ is defined as \citep{refId0}:
\begin{equation}
    \label{equ:coul}
     \begin{aligned}
L_{\text {coul }}\sim & 1.23 \times 10^{37} \operatorname{erg} \mathrm{s}^{-1}\left(\frac{\Lambda}{0.1}\right)^{-7 / 12}\left(\frac{\tau} {20}\right)^{7 / 12} \\
& \times \left(\frac{M_{*}}{1.4 M_{\odot}}\right)^{11 / 8} \left(\frac{R_{*}}{10 \mathrm{~km}}\right)^{-13 /24}\left(\frac{E_{\rm cyc}}{10 \mathrm{keV}}\right)^{-1 / 3},
\end{aligned} 
\end{equation}
where $\Lambda =1$ for spherical accretion and $\Lambda < 1$ for disk accretion \citep{1973ApJ...184..271L}, the neutron star mass $M_{*}$ of Vela X-1 $\sim 1.77 M_{\odot}$ \citep{2011ApJ...730...25R}, the typical radius of the neutron star $R_{*} = 10\ \rm km$, Thomson optical depth $\tau \sim 20$, and the CRSF energy $E_{\rm cyc}$.

In \cref{fig:GE1E2_Flux}, we plot the range of $L_{\text {coul}}$ (blue dashed lines) in the case of $\Lambda = 0.1$. One can find that most of the observed luminosity $L_{\rm X}$ is at $\lesssim L_{\rm coul} \textless L_{\rm crit}$, which means the falling materials should be stopped at the neutron star surface and thus no correlations are expected. In this case, the observed X-ray luminosity didn't reach the Coulomb interaction bound, which is necessary to interpret the positive correlation between cyclotron line energy and luminosity. \cite{2014ApJ...780..133F} has proposed a possible explanation that the reduced accretion column radius leads to a reduced Coulomb luminosity in the spherical accretion case (see their Figure 7). Considering the similar spherical accretion scenario, i.e. $\Lambda =1$, then the Coulomb luminosity for $E_{\rm cyc} = 25 \ \rm keV$ is shown in the middle panel of \cref{fig:GE1E2_Flux} with vertical magenta dashed lines. The lower limit and higher limit represent the mass of the neutron star is $1.4 M_{\odot}$ and $1.8 M_{\odot}$, respectively. The Coulomb luminosity range is about $2 \times 10^{36}\rm \ erg \ s^{-1}$, and above that luminosity a positive correlation between $E_1 - L_{\rm X}$ would be expected.


In the bottom panel of \cref{fig:GE1E2_Flux}, the relation between photon index $\Gamma$ and luminosity is also presented. There is a weak anti-correlation between the photon index and luminosity with Pearson’s correlation coefficient $r\sim -0.42$ based on the Insight-HXMT data, which is also generally consistent with the results from NuSTAR which used the same spectral model \citep{2014ApJ...780..133F}. \cite{2013ApJ...767...70O} also found that the photon index $\Gamma$ of NPEX is anti-correlated with luminosity using Suzaku data. This feature of anti-correlation between $\Gamma$ and luminosity could be typical for sub-critical accretion regime source, e.g., the other accretion pulsar Her X-1 \citep{2007A&A...465L..25S}. 

\subsection{Phase dependence of CRSF and continuum parameters}
\label{sec: phased}

\begin{figure*}
\centering
\includegraphics[width=.33\textwidth]{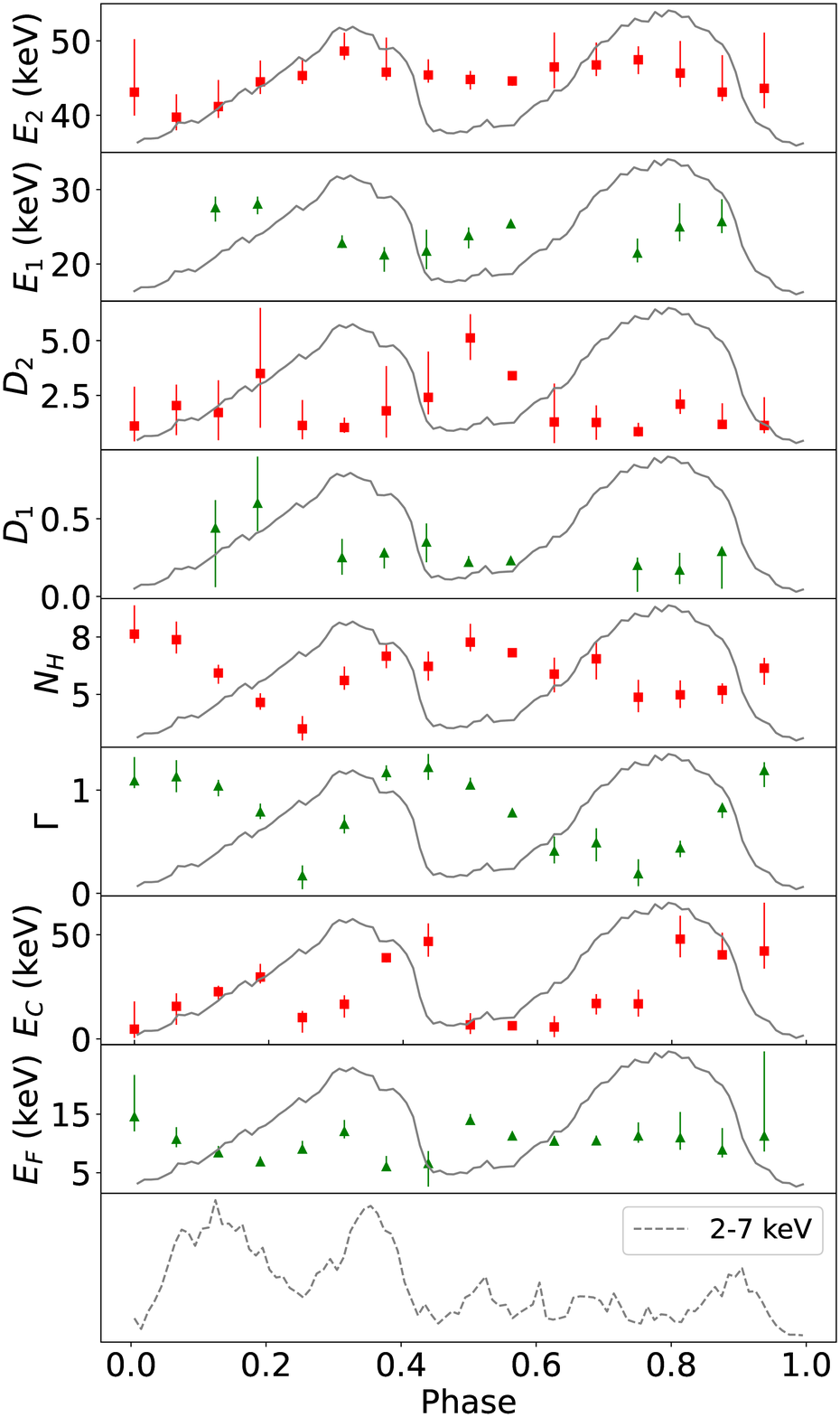}
\includegraphics[width=.33\textwidth]{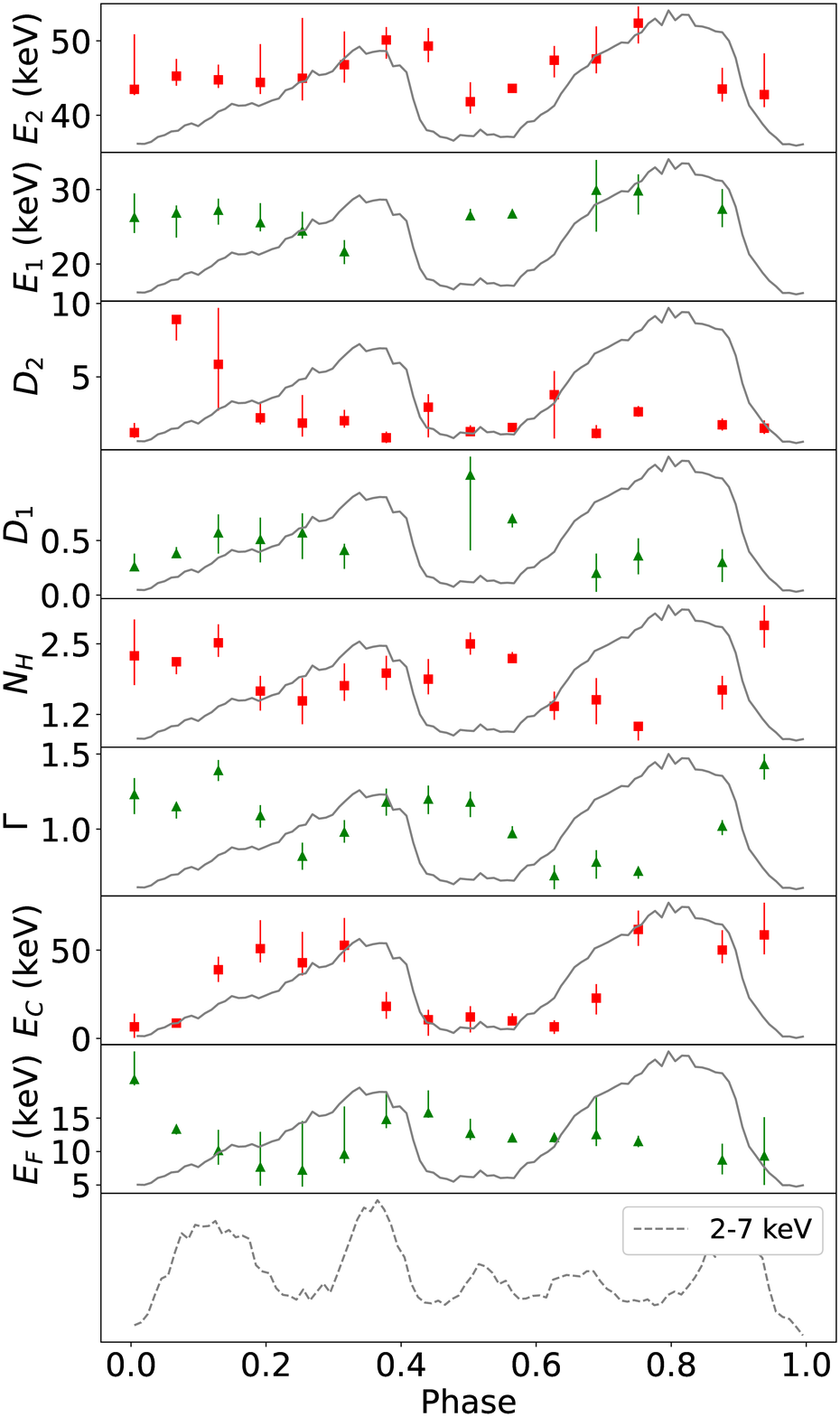}
\includegraphics[width=.33\textwidth]{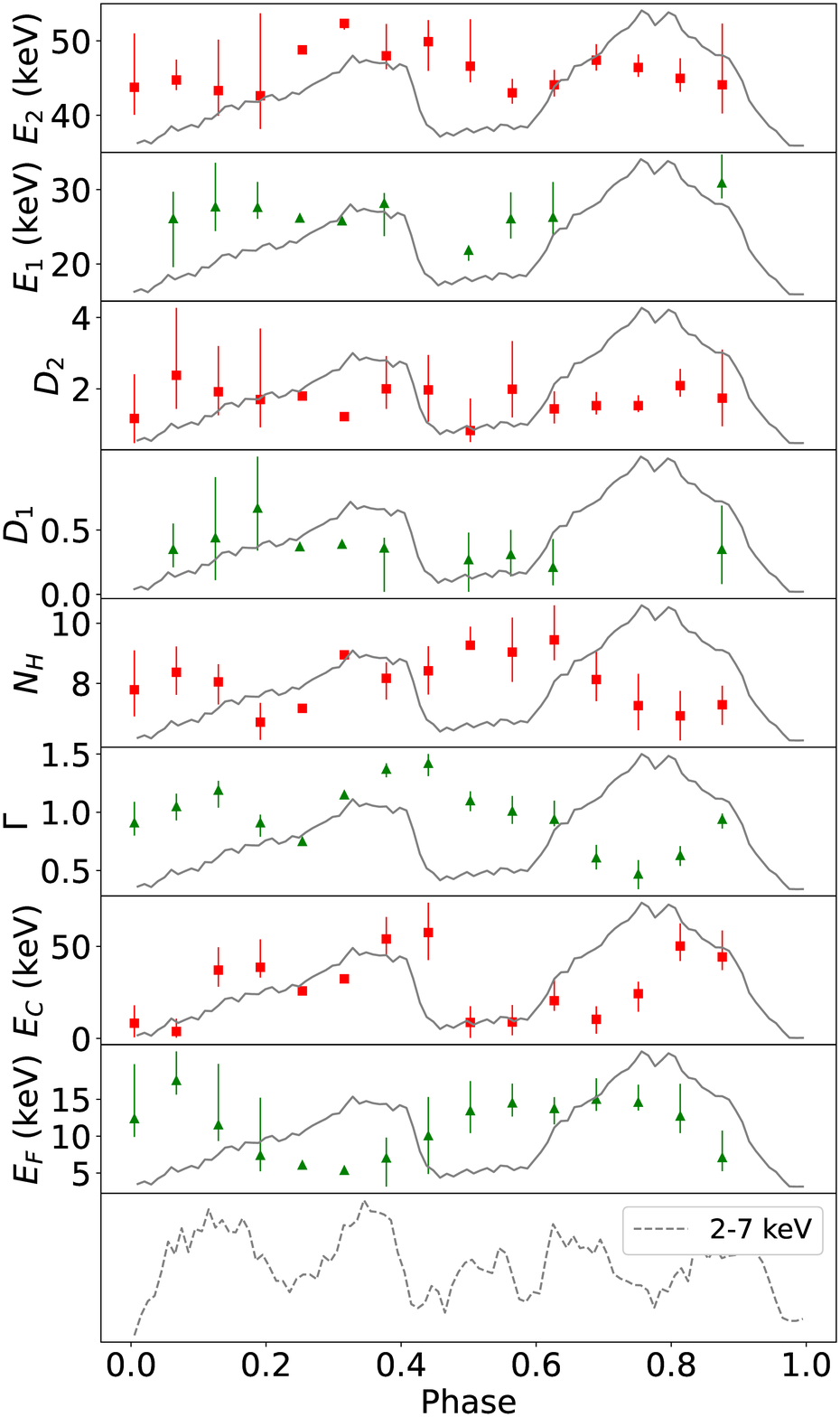}

\centerline{(a)\hspace{6cm}(b)\hspace{6cm}(c)}
\caption{The evolution of fitted cyclotron line parameters and the continuum parameters over the phase pulse in phase resolved spectroscopy for three observations: (a) ObsId: 14906 (b) ObsId: 14903 (c) ObsId: 14914. The neutral hydrogen column density $N_{\rm H}$ is in units of 10$^{22}$ atoms cm$^{-2}$. The solid gray lines in the sub panels indicate the pulse profile in the band of 25- 50 keV, while the 2 - 7 keV pulse profile (dashed gray line) is shown in the bottom panel of each observation for comparison. The missing data points in the fundamental cases illustrate that the fundamental cyclotron lines are not detected. Errors are given at the 90\% confidence.}
\label{fig:Par_Phase}
\end{figure*}

Phase-resolved spectroscopy of X-ray pulsars provides the description
of the geometric structure of accretion flow and the magnetic field around the star and consequently a deeper understanding of the accretion column physics on the surface of the neutron star \citep{2012MNRAS.420..720M}. In the previous researches, pulse phase resolved spectral analysis of the cyclotron line parameters have been studied in some X-ray pulsars such as 4U 0115+63 \citep{2000AIPC..510..173H}, Cen X-3 \citep{2008ApJ...675.1487S} and GX 301-2 \citep{2012ApJ...745..124S}. 

We have confirmed the presence of two CRSFs after performing a phase resolved analysis of Vela X-1, and presented the variation of CRSF features with the pulse phase. \cref{fig:Par_Phase} shows the variation of the best-fit CRSF parameters over the pulse phase for three representative observations. The two cyclotron line energies show different behaviors with the pulse phase. The harmonic CRSF energy commonly appears to have a positive correlation with the X-ray flux, the line energy of $\sim 50$ keV around the peak phase, and $\sim 40$ keV around the valley. This correlation between harmonic energy and the intensity of the pulsation was also confirmed by \cite{2003A&A...400..993L} with the BeppoSAX satellite. The fundamental line, however, shows a little complex trend with the pulse profile. The fundamental line energy approximates a constant or has a decline trend around the peak phase $\sim 0.3-0.4$ with a minimum value of $\sim 20$ keV. The CRSF line (specially for the weak fundamental line) is not always seen at some of pulse phases \citep{2002A&A...395..129K} due to limited statistics, thus we cannot make a comprehensive comparison between the fundamental line evolution and the pulse phase. \cite{2013ApJ...763...79M} also found a decreasing trend of $E_1$ with phase for the secondary pulse peak with nearly constant values for the main pulse. The cyclotron line absorption depths of the two lines do not change significantly with the pulse profile. The harmonic CRSF depth is around $\sim 2$, and fundamental one is around $\sim 0.3-0.5$.


In \cref{fig:Par_Phase}, the variation of continuum spectral parameters over the pulse phase is also presented. The neutral hydrogen column density $N_{\rm H}$ has a large variation over the pulse phase and it can change from $\sim (1- 10)\times 10^{22}$ atoms $\rm cm^{-2}$. $N_{\rm H}$ has a minimum at the phases $\sim 0.25$ and $\sim 0.8$ which are the peak phases of the pulse profile, and reaches a maximum at the phases $\sim 1.0$ and $\sim 0.5$ which are the valley phases. \cite{2010A&A...519A..37F} also reported a systematic change of the column density in different pulse phases which is consistent with our results. The photon index $\Gamma$ is also strongly dependent on the pulse profile, and has a similar varying curve with that of $N_{\rm H}$. The spectra became harder with ascending pulse profile peaking at $\Gamma\sim 0.5$ and softer in the declining phases ($\Gamma\sim 1.2-1.5$). The other two continuum parameters show complicated changes with the pulse phase. The variation of the cutoff energy $E_{c}$ in the three observations has a good agreement with pulse phase with the maximum at peak phase and minimum at the valley, which is similar to the report in \cite{2003A&A...400..993L}. The folding energy $E_{\rm f}$ has the limited change over the whole pulse phase with the values from $\sim 5-16$ keV. 

\begin{figure}
\centering
\includegraphics[width=.5\textwidth]{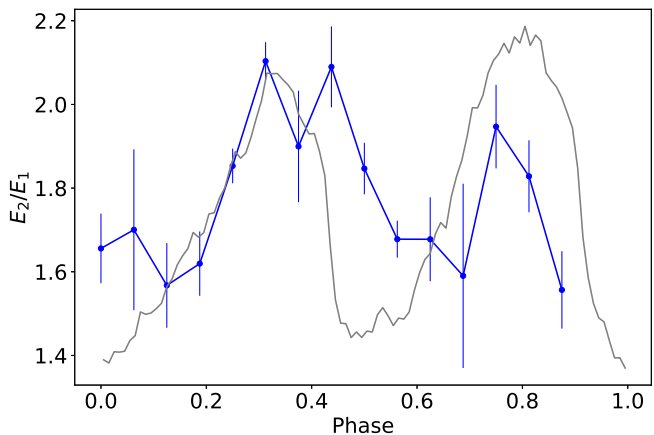}
\caption{The ratio of two CRSF line centroid energies versus pulse phase. The solid gray line indicates the pulse profile in the band of 25- 50 keV.}
\label{fig:E1_E2_Phase}
\end{figure}

The averaged phase spectral results (also see \cref{fig:Etime}) have shown that the ratio of $E_{\rm 2}/ E_{\rm 1}$ in Vela X-1 changes significantly. We also plot this ratio in different pulse phases by combining results of the phase-resolved spectroscopy work in \cref{fig:Par_Phase}. As seen in \cref{fig:E1_E2_Phase}, the line energy ratio of the harmonic line to fundamental line shows the strong variations with the pulse profile, the ratio is about $1.9-2.1$ near the peak phase, and reduces to $\sim 1.5-1.6$ around the off-peak phase. Vela X-1 shows a shallower and narrower fundamental line and a deeper and wider harmonic line for both phase-averaged and phase-resolved spectra. 

The substantial variable photoelectric absorption ($N_{\rm H} \sim (1-30)\times 10^{22} \ \rm cm^{-2}$) is observed in the phase-average spectra of Vela X-1, and also varies over different pulse phases. In \cref{fig:Par_Phase}, $N_{\rm H}$ changes from $(3-8)\times 10^{22} \rm cm^{-2}$ for the observation (a), and from $(1-3)\times 10^{22}\rm cm^{-2}$ for (b), while from $(7-10)\times 10^{22} \rm cm^{-2}$ for (c), these strong variations of column density in Vela X-1 suggest the inhomogeneous structure of winds from the supergiant donor (also see \citealt{2010A&A...519A..37F,2015MNRAS.447.1299S,2016A&A...588A.100M,2017A&A...608A.143G,2021A&A...652A..95K}). The dips of the pulsar profiles at phase 0 and 0.5 would be due to the increase of this additional absorption column density (e.g., through accretion flow matter). The changes in the value of $N_{\rm H}$ with pulse phase could be a tracer for the properties of the plasma in the accretion stream, which may due to a clumpy structure of stellar winds. The soft band pulse profiles from 2 -- 7 keV also show a dip around 0.25, where $N_{\rm H}$ is also at minimum, so that this soft band pulse dip could not be induced by the increasing absorption. The possible hot spots and heated plasma in the structured accretion flow could be the multiple soft X-ray emission regions in the magnetosphere, attributing to the multiple peaks in the soft X-ray pulse profiles. 


The centroid line energy ratio of two lines varies from $\sim 1.5 -2.1$ over the pulse phase, and the ratio follows the hard X-ray pulse profile well, which is similar to the results reported by \cite{2013ApJ...763...79M}. The cyclotron line features in Vela X-1 could be the superposition of a large number of lines from different sites or heights in the line-forming region \citep{2011ApJ...730..106N}. The fundamental line properties suggest that the viewing angle is larger, and a fan-like beam is thus expected \citep{2011ApJ...730..106N}. This scenario favors the line formation regions in the different heights: the fundamental line forms at a higher site and second harmonic primarily forms around the star surface. The recent simulations of the cyclotron absorption line characteristics over the pulse phase predict that the ratio of the harmonic line energy to the fundamental line energy changes significantly from $\sim 1.6$ to $\sim 2.2$ with viewing angles, and the fundamental line can be too shallow to be seen at some viewing angles, which is well consistent with our observational results (Nishimura 2022, submitted).

\section{Conclusion} \label{sec:summary}

In summary, we investigated the X-ray properties of Vela X-1 from 2 -- 105 keV with the multiple Insight-HXMT observations, and reported the fundamental line energy at $\sim 21-27$ keV and the first harmonic line energy varying between $\sim$ 43 -- 50 keV. Vela X-1 has the X-ray luminosity ranges of $\sim (1-20)\times 10^{36} \rm erg \ s^{-1}$ in the range of 2 -- 100 keV. The harmonic line energy has no evolution with the luminosity, while the fundamental line centroid energy shows a positive relation with X-ray luminosity, which indicates the sub-critical accreting regime in Vela X-1. Possible negative relation of $E_{\rm cyc}-L_{\rm X}$ in the low luminosity should be studied with more observations. The photon index $\Gamma$ has the weak anti-correlation with both the luminosity and fundamental line energy. In addition, the multi-band pulse profiles are shown, the multiple peak profiles below $\sim 10$ keV evolve to the double peak feature in the higher energy bands. 

Vela X-1 shows the strong variations of continuum parameters and cyclotron line features over the pulse profiles, discovering the line energy ratio of two CRSFs show the double peaks similar to the hard X-ray pulse profiles, with the values at $\sim 1.6$ at the valley phase, and $\sim 2.0$ around the peak phase. The variations of two cyclotron line features would be the results of the superposition of a large number of lines from different heights, and the line forming region of the fundamental would be at a higher site and second harmonic at the bottom. In the accretion system of Vela X-1, the absorption column density shows the long-term changes from $\sim (1-30)\times 10^{22} \rm cm^{-2}$, and significant variations over the pulse phase, implying the existence of the clumpy wind structure.

\section*{Acknowledgements}
We are grateful to the referee for the fruitful comments to improve the manuscript and thank Osamu Nishimura for the discussion and providing us the latest simulating results, and are also grateful to L.D. Kong, Y.L. Tuo for the help on the HXMT softwares. This work is supported by the National Key Research and Development Program of China (Grants No. 2021YFA0718500, 2021YFA0718503), the NSFC (12133007, U1838103, 11622326, U1838201, U1838202), and the Fundamental Research Funds for the Central Universities (No. 2042021kf0224). This work made use of data from the \textit{Insight}-HXMT mission, a project funded by China National Space Administration (CNSA) and the Chinese Academy of Sciences (CAS).

\section*{Data Availability}
Data that were used in this paper are from Institute of High Energy Physics Chinese Academy of Sciences(IHEP-CAS) and are publicly available for download from the \textit{Insight}-HXMT website. To process and fit the spectrum and obtain folded light curves, this research has made use of XRONOS and FTOOLS provided by NASA.

\bibliographystyle{mnras}
\bibliography{references}

\label{lastpage}

\begin{table*}
    \centering
    \caption{The best-fitting spectral parameters of Vela X-1 in the hard X-ray bands from 2–105 keV discussed in \cref{sec:result}. The best model is described with FD-cut multiplied by phabs (XSPEC) plus two cyclotron scattering lines, sometimes with an Fe K$\alpha$ line around 6.4 keV. The continuum flux is given in units of erg cm$^{-2}$ s$^{-1}$, and the neutral hydrogen column density in units of 10$^{22}$ atoms cm$^{-2}$. The other parameters $E_c, \ E_f, \ E1, \ E_2, \ W_1, \ W_2, \ E_{Fe}$ and $\sigma_{Fe}$ are in units of keV. Uncertainties are given at the 68\% confidence.}
    \label{tab:out_fraction}
    \renewcommand\arraystretch{1.5}
    \setlength{\tabcolsep}{0.27mm}{
    \begin{tabular}{l|ccccccccccccccc}
    \hline
     ExpID & $N_{H}$& $\Gamma$ & $E_{c}$ & $E_{f}$ & $D_2$ & $E_2$ & $W_2$ & $D_1$ & $E_1$ & $W_1$ & $E_{Fe}$ & $\sigma_{Fe}$ & Log Flux & $\chi^2$/dof\\ 
    \hline 
00201	& $9.07_{-0.34}^{+0.32}$ & $1.06_{-0.05}^{+0.05}$ & $16.39_{-3.85}^{+3.24}$ & $13.36_{-0.86}^{+1.01}$ & $1.04_{-0.28}^{+0.61}$ & $46.95_{-1.16}^{+1.65}$ & $3.28_{-1.49}^{+3.51}$ 	& & & & & & $-8.393_{-0.004}^{+0.004}$ 	& 999 /1237	\\ \hline
00203	& $21.49_{-0.88}^{+0.80}$ & $1.17_{-0.06}^{+0.05}$ & $12.42_{-5.78}^{+4.59}$ & $18.24_{-0.82}^{+0.92}$ & $1.98_{-0.53}^{+0.83}$ & $46.24_{-0.56}^{+0.55}$ & $2.23_{-0.63}^{+0.92}$ 	& & & & & & $-8.490_{-0.005}^{+0.004}$	& 842 /1259	\\ \hline
00301	& $25.00_{-1.19}^{+1.26}$ & $0.96_{-0.08}^{+0.09}$ & $13.35_{-5.76}^{+6.40}$ & $15.76_{-0.91}^{+0.88}$ & $1.60_{-0.35}^{+0.73}$ & $48.30_{-0.72}^{+0.80}$ & $3.65_{-1.34}^{+1.41}$ & $0.12_{-0.06}^{+0.08}$ & $25.15_{-1.15}^{+1.07}$ & $1.99_{-0.78}^{+2.22}$ 	& & & $-8.448_{-0.004}^{+0.004}$	& 914/1324	\\ \hline
14106	& $16.42_{-0.61}^{+0.63}$ & $1.41_{-0.05}^{+0.05}$ & $36.65_{-3.38}^{+3.98}$ & $12.75_{-1.10}^{+1.17}$ & $1.01_{-0.15}^{+0.22}$ & $46.61_{-1.03}^{+1.17}$ & $5.63_{-2.21}^{+2.86}$ 	& & & & & & $-8.151_{-0.004}^{+0.004}$	& 818 /880	\\ \hline
14301	& $9.76_{-0.28}^{+0.27}$ & $1.08_{-0.04}^{+0.05}$ & $9.98_{-4.16}^{+4.10}$ & $15.09_{-0.82}^{+0.84}$ & $1.37_{-0.20}^{+0.30}$ & $48.95_{-1.07}^{+1.37}$ & $5.32_{-1.58}^{+2.02}$ 	& & & & & & $-8.258_{-0.002}^{+0.003}$	& 1077/1344	\\ \hline
14302	& $12.91_{-0.43}^{+0.47}$ & $0.93_{-0.05}^{+0.05}$ & $16.65_{-5.49}^{+3.72}$ & $13.47_{-1.50}^{+2.42}$ & $1.24_{-0.26}^{+0.41}$ & $50.02_{-1.64}^{+2.36}$ & $10.59_{-4.09}^{+6.16}$ 	& & & & & & $-8.096_{-0.003}^{+0.003}$	& 958 /1151	\\ \hline
14401	& $19.62_{-0.51}^{+0.47}$ & $1.26_{-0.04}^{+0.04}$ & $30.43_{-2.49}^{+2.97}$ & $12.17_{-0.81}^{+0.89}$ & $0.93_{-0.12}^{+0.12}$ & $45.35_{-0.79}^{+0.93}$ & $6.10_{-1.77}^{+2.38}$ & $0.17_{-0.04}^{+0.03}$ & $23.14_{-0.91}^{+0.74}$ & $6.19_{-1.95}^{+3.01}$ 	& & & $-8.139_{-0.003}^{+0.003}$	& 964/1192	\\ \hline
14402	& $20.43_{-0.61}^{+0.62}$ & $1.08_{-0.06}^{+0.04}$ & $17.35_{-4.00}^{+3.27}$ & $14.44_{-0.66}^{+0.87}$ & $1.12_{-0.20}^{+0.27}$ & $46.13_{-0.71}^{+0.89}$ & $4.16_{-1.38}^{+1.69}$ & $0.11_{-0.04}^{+0.05}$ & $23.76_{-1.05}^{+1.02}$ & $2.90_{-1.37}^{+2.40}$ 	& & & $-8.230_{-0.004}^{+0.004}$	& 940/1249	\\ \hline
14405	& $12.73_{-0.34}^{+0.30}$ & $1.03_{-0.08}^{+0.05}$ & $21.42_{-2.54}^{+2.59}$ & $10.63_{-0.84}^{+0.67}$ & $1.31_{-0.35}^{+0.42}$ & $43.35_{-0.82}^{+1.06}$ & $1.91_{-0.51}^{+0.71}$ & $0.18_{-0.04}^{+0.04}$ & $24.75_{-1.52}^{+1.55}$ & $4.95_{-1.90}^{+3.01}$	& & & $-8.237_{-0.005}^{+0.005}$	& 940/1145	\\ \hline
14501	& $9.51_{-0.33}^{+0.35}$ & $1.00_{-0.05}^{+0.05}$ & $18.01_{-3.43}^{+3.38}$ & $12.58_{-0.55}^{+0.51}$ & $1.20_{-0.24}^{+0.34}$ & $46.70_{-0.68}^{+0.80}$ & $3.01_{-0.83}^{+1.27}$ & $0.15_{-0.04}^{+0.04}$ & $23.61_{-0.82}^{+0.97}$ & $3.83_{-1.46}^{+2.73}$ 	& & & $-8.093_{-0.003}^{+0.003}$	& 898/1038	\\ \hline
14503	& $15.39_{-0.57}^{+0.54}$ & $1.18_{-0.05}^{+0.05}$ & $26.07_{-5.19}^{+4.27}$ & $15.03_{-1.28}^{+1.71}$ & $1.11_{-0.16}^{+0.18}$ & $44.98_{-1.22}^{+1.19}$ & $8.96_{-3.17}^{+5.51}$ & $0.25_{-0.07}^{+0.06}$ & $22.61_{-0.67}^{+0.88}$ & $2.91_{-1.20}^{+1.57}$ 	& & & $-8.296_{-0.006}^{+0.006}$	& 759/957	\\ \hline
14504	& $15.08_{-0.51}^{+0.50}$ & $0.97_{-0.05}^{+0.06}$ & $18.56_{-2.16}^{+2.16}$ & $10.78_{-0.45}^{+0.54}$ & $1.59_{-0.58}^{+1.04}$ & $49.00_{-1.25}^{+1.57}$ & $1.72_{-0.55}^{+1.31}$ 	& & & & & & $-8.151_{-0.002}^{+0.002}$	& 1072/1258	\\ \hline
14508	& $16.99_{-0.45}^{+0.57}$ & $1.38_{-0.04}^{+0.05}$ & $19.72_{-3.95}^{+4.57}$ & $18.58_{-1.28}^{+0.89}$ & $1.40_{-0.28}^{+0.45}$ & $45.52_{-0.78}^{+1.07}$ & $2.27_{-0.62}^{+0.72}$ & $0.16_{-0.03}^{+0.04}$ & $21.58_{-1.29}^{+0.90}$ & $6.01_{-2.27}^{+3.59}$ 	& & & $-8.279_{-0.005}^{+0.005}$	& 1342/1425	\\ \hline
14703	& $33.30_{-1.49}^{+1.51}$ & $0.88_{-0.08}^{+0.07}$ & $23.41_{-2.18}^{+1.98}$ & $9.86_{-0.49}^{+0.52}$ & $1.10_{-0.42}^{+0.58}$ & $46.34_{-1.13}^{+1.76}$ & $1.59_{-0.46}^{+1.87}$ 	& & & & & & $-8.139_{-0.004}^{+0.003}$	& 812/890	\\ \hline
14704	& $23.88_{-0.69}^{+0.62}$ & $0.93_{-0.05}^{+0.04}$ & $31.08_{-2.36}^{+2.50}$ & $9.64_{-0.57}^{+0.65}$ & $0.80_{-0.15}^{+0.15}$ & $45.30_{-0.80}^{+0.84}$ & $9.15_{-2.54}^{+2.80}$ & $0.16_{-0.04}^{+0.04}$ & $24.67_{-0.93}^{+1.06}$ & $7.19_{-2.18}^{+3.11}$ 	& & & $-7.946_{-0.002}^{+0.002}$	& 1046/1170	\\ \hline
14705	& $16.37_{-0.47}^{+0.45}$ & $0.82_{-0.04}^{+0.04}$ & $25.19_{-1.03}^{+1.03}$ & $9.04_{-0.29}^{+0.31}$ & $0.53_{-0.16}^{+0.25}$ & $44.84_{-0.90}^{+1.29}$ & $1.91_{-0.71}^{+2.23}$ 	& & & & $6.47_{-0.03}^{+0.04}$ & $0.04_{-0.04}^{+0.07}$ & $-7.891_{-0.002}^{+0.002}$	& 1072/1188	\\ \hline
14707	& $16.37_{-0.61}^{+0.66}$ & $0.73_{-0.06}^{+0.06}$ & $8.70_{-6.07}^{+5.19}$ & $12.75_{-1.42}^{+3.24}$ & $0.67_{-0.17}^{+0.30}$ & $44.66_{-1.91}^{+2.07}$ & $11.27_{-7.14}^{+14.06}$ & & &	& & & $-8.028_{-0.004}^{+0.004}$	& 1300/1343	\\ \hline
14709	& $19.94_{-0.85}^{+0.89}$ & $1.04_{-0.07}^{+0.06}$ & $21.38_{-3.90}^{+3.28}$ & $12.98_{-1.10}^{+1.41}$ & $0.80_{-0.16}^{+0.21}$ & $46.74_{-1.45}^{+1.70}$ & $9.81_{-4.15}^{+6.40}$ 	& & & & & & $-8.146_{-0.004}^{+0.004}$	& 797 /858	\\ \hline
14901	& $2.11_{-0.08}^{+0.07}$ & $1.02_{-0.02}^{+0.02}$ & $38.87_{-2.71}^{+2.96}$ & $8.89_{-0.77}^{+0.91}$ & $1.24_{-0.22}^{+0.27}$ & $45.44_{-1.19}^{+1.21}$ & $16.65_{-3.46}^{+3.67}$ & $0.24_{-0.05}^{+0.05}$ & $26.39_{-0.51}^{+0.56}$ & $6.33_{-1.14}^{+1.24}$ & $6.53_{-0.04}^ {+0.06}$ & $0.15_{-0.08}^{+0.11}$ & $-7.762_{-0.002}^{+0.002}$	& 1162/1336	\\ \hline
14903	& $1.99_{-0.09}^{+0.08}$ & $1.12_{-0.02}^{+0.02}$ & $40.15_{-3.48}^{+3.47}$ & $9.42_{-1.00}^{+1.06}$ & $1.24_{-0.20}^{+0.19}$ & $46.16_{-1.09}^{+1.16}$ & $10.39_{-2.87}^{+3.53}$ & $0.33_{-0.07}^{+0.09}$ & $26.40_{-0.89}^{+1.05}$ & $8.04_{-2.01}^{+2.62}$ 	& & & $-7.912_{-0.003}^{+0.002}$	& 1176/1326	\\ \hline
14906	& $6.19_{-0.16}^{+0.16}$ & $0.94_{-0.02}^{+0.02}$ & $36.43_{-1.59}^{+1.94}$ & $7.57_{-0.52}^{+0.51}$ & $0.87_{-0.13}^{+0.15}$ & $44.52_{-0.71}^{+0.71}$ & $6.27_{-1.83}^{+2.45}$ & $0.36_{-0.05}^{+0.06}$ & $25.70_{-0.70}^{+0.89}$ & $9.75_{-1.70}^{+2.19}$ & $6.53_{-0.06}^{+0.07}$ & $0.09_{-0.09}^{+0.13}$ & $-7.753_{-0.002}^{+0.002}$	& 1081/1226	\\ \hline
14910	& $5.17_{-0.13}^{+0.13}$ & $1.09_{-0.02}^{+0.02}$ & $30.78_{-1.10}^{+1.28}$ & $9.51_{-0.70}^{+0.73}$ & $0.74_{-0.15}^{+0.20}$ & $44.64_{-1.09}^{+1.11}$ & $6.24_{-3.27}^{+3.72}$ 	& & & & & & $-8.033_{-0.002}^{+0.002}$	& 1184/1376	\\ \hline
14911	& $4.42_{-0.15}^{+0.19}$ & $0.89_{-0.03}^{+0.04}$ & $27.00_{-1.14}^{+1.41}$ & $8.80_{-0.69}^{+2.49}$ & $1.12_{-0.42}^{+0.66}$ & $48.75_{-1.54}^{+2.85}$ & $5.07_{-3.26}^{+12.56}$ & & & & & & $-8.017_{-0.002}^{+0.002}$	& 1272/1269	\\ \hline
14913	& $7.67_{-0.26}^{+0.25}$ & $1.04_{-0.03}^{+0.04}$ & $28.73_{-0.80}^{+0.91}$ & $7.44_{-0.50}^{+0.60}$ & $0.68_{-0.41}^{+0.75}$ & $46.50_{-1.91}^{+3.95}$ & $1.99_{-0.77}^{+3.02}$ 	& & & & & & $-7.971_{-0.003}^{+0.003}$	& 1069/1118	\\ \hline
14915	& $7.90_{-0.28}^{+0.25}$ & $0.93_{-0.03}^{+0.03}$ & $62.44_{-5.61}^{+4.32}$ & $3.38_{-1.80}^{+1.57}$ & $2.74_{-0.23}^{+0.15}$ & $46.43_{-1.01}^{+1.15}$ & $15.77_{-3.56}^{+3.39}$ & $0.37_{-0.11}^{+0.16}$ & $26.43_{-0.88}^{+1.14}$ & $8.97_{-2.34}^{+3.01}$ 	& & & $-7.568_{-0.003}^{+0.003}$	& 1060/1103	\\ \hline
15002	& $31.65_{-1.14}^{+1.32}$ & $1.32_{-0.06}^{+0.06}$ & $30.59_{-3.96}^{+3.61}$ & $13.92_{-1.15}^{+1.38}$ & $1.47_{-0.28}^{+0.46}$ & $47.98_{-1.09}^{+1.12}$ & $5.58_{-2.05}^{+2.90}$ 	& & & & & & $-8.049_{-0.004}^{+0.005}$	& 644/801	\\ \hline
15103	& $7.63_{-0.11}^{+0.12}$ & $0.94_{-0.02}^{+0.02}$ & $49.34_{-2.58}^{+2.96}$ & $9.42_{-0.85}^{+0.94}$ & $2.21_{-0.17}^{+0.16}$ & $47.09_{-0.65}^{+0.70}$ & $19.37_{-1.80}^{+1.98}$ & $0.30_{-0.04}^{+0.04}$ & $26.45_{-0.30}^{+0.35}$ & $7.53_{-0.95}^{+1.06}$ 	& & & $-7.419_{-0.001}^{+0.001}$	& 1369/1375	\\ \hline
15203	& $15.94_{-0.66}^{+0.49}$ & $1.38_{-0.05}^{+0.05}$ & $49.87_{-2.56}^{+2.11}$ & $8.02_{-1.08}^{+1.18}$ & $3.56_{-0.96}^{+1.16}$ & $44.24_{-0.45}^{+0.43}$ & $2.37_{-0.61}^{+0.76}$ & $0.43_{-0.06}^{+0.06}$ & $24.41_{-0.89}^{+0.76}$ & $8.78_{-1.57}^{+1.68}$ 	& & & $-8.420_{-0.007}^{+0.007}$	& 1244/1341	\\ \hline
	
    \end{tabular} 
    }
\end{table*}

\end{document}